\documentclass[%
 aip,
 amsmath,amssymb,
preprint,%
]{revtex4-1}

\usepackage{graphicx}
\usepackage{dcolumn}
\usepackage{bm}

\usepackage[utf8]{inputenc}
\usepackage[T1]{fontenc}
\usepackage{mathptmx}
\usepackage{makecell}
\usepackage{xcolor}

\usepackage{amsmath}
\usepackage{amsfonts}
\usepackage{amssymb}
\usepackage{natbib}

\newcommand{\velamp}{\boldsymbol{\mathsf{U}}}
\newcommand{\velocityeigenvalue}{\boldsymbol{\lambda}_{\velamp}}

\newcommand{\temperatureeigenvalueconstantsolid}{\boldsymbol{\lambda}_{\theta}}
\newcommand{\temperatureeigenvaluequasideveloped}{\boldsymbol{\lambda}_{\Theta}}

\begin{document}


\title[]{A Mathematical Description of the Quasi-Periodically Developed Heat Transfer Regime in Channels with Arrays of Periodic Solid Structures}

\author{G. Buckinx}%
\email{geert.buckinx@vito.be.}
\affiliation{%
VITO, Boeretang 200, 2400 Mol, Belgium
}%
\affiliation{%
EnergyVille, Thor Park, 3600 Genk, Belgium
}%

\date{\today}

\begin{abstract}
In this article, we present the governing equations for the temperature field upstream and downstream of the periodically developed flow region in channels with arrays of periodic solid structures.
From the ansatz that the temperature field in this region is determined by exponentially decaying modes, just like the quasi-developed flow field, we arrive at an eigenvalue problem which governs the allowed temperature modes.
This eigenvalue problem is virtually identical to the one for periodically developed heat transfer in isothermal solids, except that
the conjugate heat transfer in the solid is effectively taken into account.

\end{abstract}

\maketitle
\newpage

\section{Introduction}

Laminar developed flow and heat transfer in channels with a constant cross section have been well studied in the literature.
The notion of \textit{developed flow} refers to the observation that certain flow features in a channel with a constant cross section typically become invariant after a certain distance from the channel inlet.
In particular, the velocity profile and pressure gradient no longer evolve along the main flow direction, once the flow is developed.
Likewise, also the temperature field in such a developed flow will evolve towards a certain profile whose shape is similar at any cross section perpendicular to the main flow direction.
In that case one speaks of a \textit{developed heat transfer regime} in the channel.
A necessary condition for this regime to occur, is that the channel walls are either subject to a constant temperature or a constant heat flux, although some other specific types of boundary conditions are allowed as well.

For channels with a circular cross section, the developed flow solution corresponds to a parabolic  velocity profile, as known since the works of Hagen and Poiseuille \citep{SuteraSkalak1993}. 
Developed flow solutions for channels with a rectangular, triangular, or elleptical cross section were presented already near the end of the 19th century by \citet{Boussinesq1868} and \citet{Proudman1914}.
The first solutions for the developed temperature profiles in laminar channel flows were derived soon after in the works by \citet{Nusselt1910} and \citet{Graetz1882, Graetz1885}, who considered the heat transfer in a circular pipe at a constant wall temperature.
Since these seminal works, a great body of works have treated the developed flow and heat transfer regimes for different fluids and channel geometries.
A systematic overview of many analytical results on the topic can be found in the somewhat outdated, but still very relevant book by \citet{ShahLondon1978}.

Next to the developed regimes, various researchers have investigated the developing flow and temperature field near the inlet of channels.
Most of the available work has focused on  (two-dimensional) developing flows in channels with a constant cross section, like parallel-plate channels and circular pipes.
Hereto several mathematical approaches have been employed.
For instance, \citet{Chen1973} adopted an integral method, in which the channel is divided into a boundary layer near the channel wall and an inviscid core.
That way, boundary-layer theory \citep{Schlichting1997} could be applied to solve the developing flow profile.
A second method relies on a linearization of the  Navier-Stokes equations, as first explored by \citet{Langhaar1942} and \citet{Sparrow1964}. 
In this method, the advection by transversal velocity components is usually neglected, so that the linearized problem can be solved analytically in the form of an infinite series.
Another approach makes use of matched asymptotic expansions to determine first the flow solution immediately upstream of the developed region, which we call the \textit{quasi-developed} flow field.
This quasi-developed solution is then coupled to a second solution for the perturbed boundary-layer flow equations which apply near the channel inlet \citep{Schlichting1934, Kapila1973}.
A distinctive feature of quasi-developed flow is the appearance of exponentially decaying velocity and pressure modes along the main flow direction, which are governed by an eigenvalue problem similar to the Orr-Sommerfeld equation.
These modes have been thoroughly investigated for quasi-developed flow between parallel plates \citep{Nachtsheim1964, Wilson1969, Brandt1969, Sadri1997, Sadri2002}.
Experimental evidence of their existence was further provided by \citet{Asai2004}.
Yet, detailed experimental work on laminar flow development in channels has remained rather limited \citep{Ahmad2010}. 
A more common approach to analyze developing flow is through direct numerical simulation of the entire channel flow, from the inlet to the outlet.
This approach has been adopted by e.g. \citet{Gessner1981, Patankar1983, Sadri1997, Ferreira2021}.

The first solutions for the developing temperature field in channels of a constant cross section can be traced back to the seminal works of \citet{Nusselt1910} and \citet{Graetz1882, Graetz1885}.
These solutions are valid under the assumption that the flow field is laminar and (practically) fully developed, while the channel wall is subject to a  constant temperature, a constant heat flux, or a temperature proportional to heat flux.
Therefore, the problem of determining the developing temperature field for a developed flow is still  called the \textit{Graetz-Nusselt problem}.
As the Graetz-Nusselt problem was originally solved by means of separation of variables and Sturm-Liouville
theory, the original series solutions for the temperature field necessitated the evaluation of many terms to obtain a good accuracy near the channel inlet.
For that reason, \citet{Leveque1928} introduced an additional similarity transformation for the temperature field, which we now know as the \textit{Lévêque approach}.
Over the years, many variants of the classical Graetz-Nusselt problem have been studied.
Papoutsakis et al. were the first to study its extensions by including the effect of axial conduction in the fluid, as well as conjugate heat transfer within the solid \citep{Papoutsakis1980,Papoutsakis1981}.
Their series solution for the temperature field was constructed in terms of the eigenfunctions of a self-adjoint operator for the Graetz-Nusselt problem, after introducing a suitable change of variables for the temperature.
This formalism based on a self-adjoint operator was further applied by Weigand et al. to solve the
extended Graetz problem for additional boundary conditions, as well as turbulent flows, taking axial wall conduction into account \citep{Weigand1996, Weigand2001, Weigand2004, Weigand2007}.
More recent works have addressed generalizations of the Graetz-Nusselt problem towards non-axisymmetric geometries and arbitrary velocity profiles on both infinite and finite domains \citep{Pierre2009, Fehrenbach2012}.
Apart from the preceding analytical approaches, also
numerical approaches based on finite-difference, finite-element or finite-volume discretisation have been employed to study the developing temperature field in channels.
These numerical methods have been pioneered by e.g. \citet{Hwang1964, Grigull1965, Montgomery1966, Montgomery1968}.
They have been the most effective means so far to study simultaneously developing flow and heat transfer \citep{Nguyen1992, Nguyen1993}.
Nevertheless, significant experimental work on the developing heat transfer regimes has been performed today \citep{Barozzi1984, Everts2020, Everts2023}.

A generalization of the former developed flow and heat transfer regimes are the so-called \textit{periodically developed} flow and heat transfer regimes, which arise in channels whose cross section varies periodically along the main flow direction.
Such channels are omnipresent in compact heat transfer devices, which typically contain arrays of hundreds of solid structures in a periodic configuration to enhance the heat transfer \citep{KaysLondon1984, ShahLondon1978}.
Besides, they have seen recent application in microfluidic devices and ordered microporous materials, which often operate under laminar flow conditions \citep{KosarMishraPeles2005,KosarPeles2007, SiuHoQuPfefferkorn2007,MohammadiKosar2018,Zargartalebi2020}.
The first mathematical descriptions of the periodically developed regimes were formulated by \citet{Patankar1977, Patankar1978}.
According to these descriptions, periodically developed flow is driven by a constant mean pressure gradient and a periodic pressure field.
Furthermore, the periodically developed temperature field has a periodic-linear profile in the case of a periodic heat flux, and a periodic-exponential profile in the case of a constant solid temperature.
In our previous works, we have shown that the constant pressure gradient in the periodically developed regime
can also be defined as the gradient of a macro-scale pressure, which is obtained by repeated volume-averaging of the original pressure field \citep{BuckinxBaelmans2015}.
That way, there exists a consistent link with the volume-averaged models for flow through porous media
\citep{Whitaker1986, Whitaker1996}.
In addition, we have shown that Patankar's description of the developed heat transfer regime in isothermal solids can be reformulated as a periodic eigenvalue problem.
The smallest eigenvalue solving this eigenproblem corresponds to the decay rate of the true macro-scale temperature in the channel, which can again be recovered by repeated volume-averaging of the original temperature field \citep{BuckinxBaelmans2015b, BuckinxBaelmans2016}.
Therefore, we can obtain theoretical consistency with the volume-averaged models for heat transfer in porous media \citep{Quintard1997, Quintard2000}.

Whereas the literature on periodically developed flow and heat transfer is well established, very limited theoretical work seems to be available on the developing regimes in channels with arrays of periodic solid structures.
As a matter of fact, we are only aware of our previous works on the quasi-periodically developed flow regime \citep{Buckinx2022arxiv, Buckinx2023arxiv, Vangeffelen2023}.
This regime is a generalization of the quasi-developed regime in channels with a constant cross section, as it is characterized by streamwise periodic velocity and pressure modes decaying exponentially along the main flow direction.
The occurrence of this regime was recently proven for Stokes flow in the work by \citet{FepponHAL2024}, who recognized these modes as \textit{boundary-layer terms}, and connected their description to formal homogenization methods for periodic media based on two-scale asymptotics.

Given the former state of the art, the present work aims to give a first mathematical description of the developing temperature field in the region of quasi-periodically developed flow.
The remainder of this work is organized as follows.
We first set out the general problem statement in section 2.
Subsequently, in section 3, we extend the known descriptions for the periodically developed heat regime transfer by including the conjugate heat transfer in the solid.
In section 4, we propose the governing equations for the temperature field in the quasi-periodically developed regime under a constant wall temperature, as well as a constant heat flux at the channel walls.
Finally, we comment on the significance of these equations, particular for the macro-scale modelling approaches presented by \citet{BuckinxBaelmans2015b, BuckinxBaelmans2016}.

\section{Problem Statement}

\begin{figure}
	\begin{center}
		\includegraphics[scale=0.25]{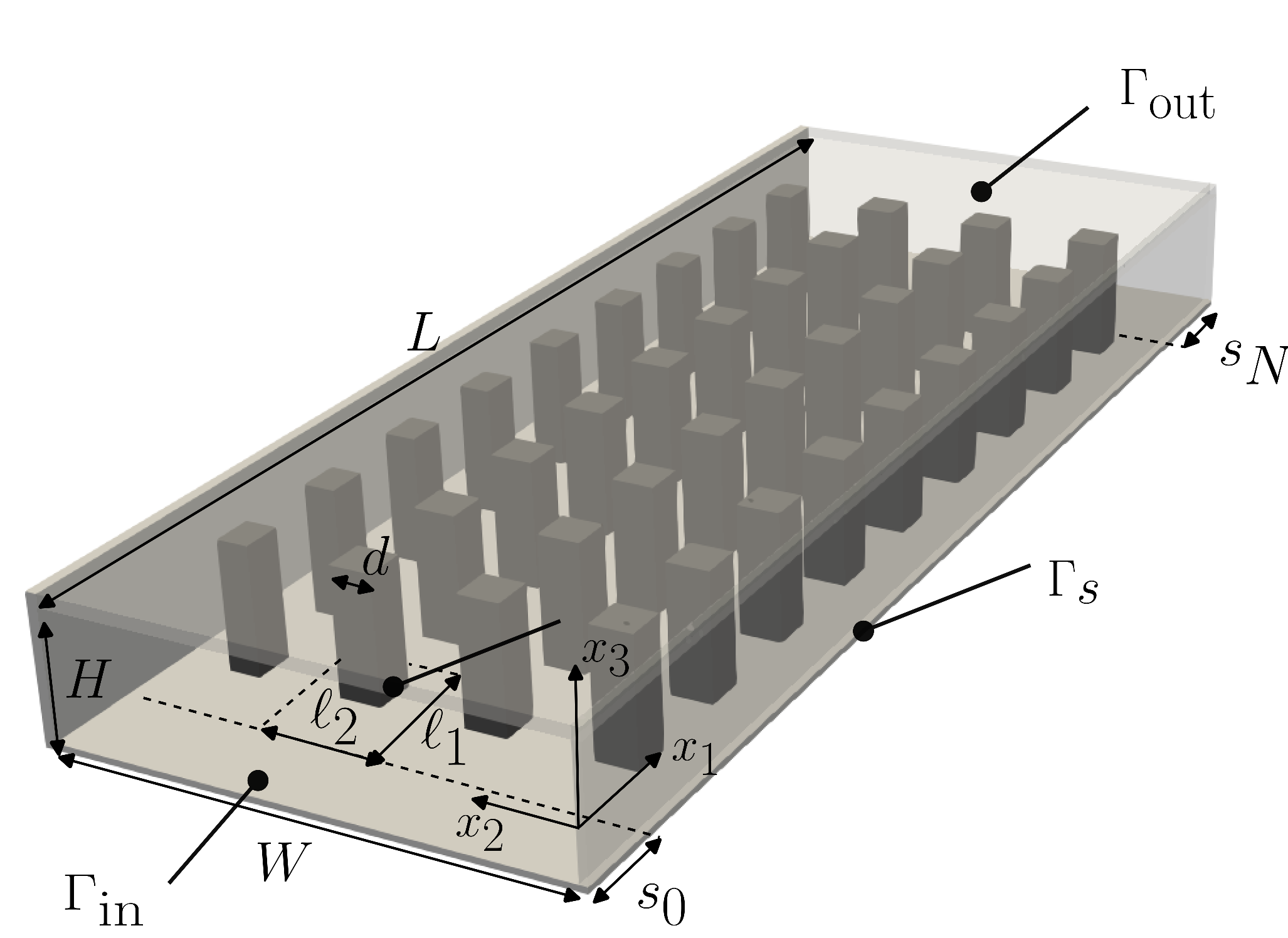}
		\caption{A channel consisting of $N_1 \times N_2$ identical unit cells, each containing a solid square cylinder of diameter $d$, at streamwise distance $\ell_1$ and a transversal distance distance $\ell_2$ from each other.}
		\label{fig: Channel and array geometry}
	\end{center}
\end{figure}

We consider a channel $\Omega$ composed of an array of $N_1 \times N_2$ identical units of fluid material ($\Omega_f$) and solid material ($\Omega_s$) separated by a fluid-solid interface $\Gamma_{fs}$, as shown in Figure \ref{fig: Channel and array geometry}.
Because the solid material distribution represented by the indicator function $\gamma_s(\boldsymbol{x})$ is spatially periodic within the array, we have at any position $\boldsymbol{x}$ sufficiently far from the channel boundaries
\begin{align}
	\gamma_s(\boldsymbol{x}) &= \gamma_s\left(\boldsymbol{x} + \boldsymbol{l}_{j} \right)
	& \mbox{for}~~ j \in \left\{1,2\right\}\,,
	\label{eq: periodic solid material distribution}
\end{align}
where $\gamma_s(\boldsymbol{x}) = 1 \leftrightarrow \boldsymbol{x}\in \Omega_s$ and $\gamma_s(\boldsymbol{x}) = 0 \leftrightarrow \boldsymbol{x}\notin \Omega_s$.
The lattice vectors $\boldsymbol{l}_{j} \triangleq \ell_j \boldsymbol{e}_j$ are by definition parallel to the unit vectors $\boldsymbol{e}_1$ and $\boldsymbol{e}_2$ along the main flow direction and transversal direction, respectively.
So, if we make use of the coordinate frame in Figure \ref{fig: Channel and array geometry}, a row of the array corresponds to all positions $x_1 \triangleq \boldsymbol{x} \boldsymbol{\cdot} \boldsymbol{e}_1 \in \left[s_0 + i\ell_1, s_0+(i-1)\ell_1 \right]$ for some integer $i \in (0, N_1)$.
Here, $s_0$ denotes the distance between the array of unit cells and the channel inlet according to Figure \ref{fig: Channel and array geometry}.
Similarily, $s_N$ denotes the distance between the array and channel outlet. 

To determine the temperature distribution $T$ in the channel, we rely on the following energy conservation equations for the fluid and solid material: 
\begin{subequations}
\label{eq: problem statement: temperature field channel flow}
\begin{align}
\rho_{f} c_{f}	\boldsymbol{u}_{f} \boldsymbol{\cdot}   \boldsymbol{\nabla} T_{f} & = 
k_{f} \nabla^2  T_{f} & \mbox{in}  ~~ \Omega_{f} \,, \\
0 &= k_s \nabla^2  T_{s}  &\mbox{in}  ~~ \Omega_{s} \,.
\end{align}
\end{subequations}
The temperature distribution is defined here as $T=T_f$ in $\Omega_f$ and $T=T_s$ in $\Omega_s$, while the velocity distribution of the fluid in the channel is defined as $\boldsymbol{u}=\boldsymbol{u}_{f}$ in $\Omega_f$ and $\boldsymbol{u}=0$ in $\Omega_s$.
The constant material properties $\rho_f$, $c_f$ and $k_f$ denote the density, heat capacity and thermal conductivity of the fluid, while $k_s$ denotes the thermal conductivity of the solid.

The temperature distribution in the channel is assumed to satisfy the following boundary conditions at the  channel boundary $\Gamma \triangleq \partial \Omega$. 
At the channel inlet $\Gamma_{\text{in}}$, the inlet 
temperature profile $T_{\text{in}}$ is given. 
At the channel outlet $\Gamma_{\text{out}}$, a zero-heat-flux condition is presumed.
For the fluid, we have thus
\begin{subequations}
\begin{align}
T_{f} \left( \boldsymbol{x} \right) &= T_{\text{in}}\left( \boldsymbol{x} \right)  
&& \mbox{on}~~\Gamma_{ \text{in}} \,, \\ 
- \boldsymbol{n} \cdot k_{f} \nabla T_{f} &= 0 
&& \mbox{on}~~\Gamma_{ \text{out}} \,,
\end{align}
\label{eq: inlet outlet bcs temperature}
\end{subequations} 
where the normal vector $\boldsymbol{n}$ at the boundary $\Gamma$ points outward of $\Omega$.
At the remaining channel boundaries formed by solid walls, i.e. $\Gamma_{s} \triangleq \Gamma \setminus \left(\Gamma_{ \text{in}} \cup \Gamma_{ \text{out}} \right)$, we consider two sets of boundary conditions. 
The first set corresponds to the case where part of the channel walls are maintained at a constant solid temperature $\bar{T}_s$, while the remaining part of the channel walls is thermally insulated:
\begin{subequations}
\begin{align}
T_{s} \left( \boldsymbol{x} \right) &= \bar{T}_s &&\mbox{on}~~\Gamma_{ \text{D}} \subset \Gamma_{s} \,, \\ 
- \boldsymbol{n} \cdot k_{s} \nabla T_{s} &= 0 && \mbox{on}~~\Gamma_{s}  \setminus \Gamma_{ \text{D}}  \,.
\end{align}
\label{eq: constant temperature and insulation walls}
\end{subequations} 
The second set of boundary conditions corresponds to the case where there is a constant heat flux $q_s$ at a certain part of the channel walls, while the remaining part of the channel walls is thermally insulated: 
\begin{subequations}
\begin{align}
- \boldsymbol{n} \cdot k_{s} \nabla T_{s}  &= q_s &&\mbox{on}~~\Gamma_{ \text{N}} \subset \Gamma_{s} \,, \\ 
- \boldsymbol{n} \cdot k_{s} \nabla T_{s} &= 0 && \mbox{on}~~\Gamma_{s}  \setminus \Gamma_{ \text{N}}  \,.
\end{align}
\label{eq: constant heat flux and insulation walls}
\end{subequations} 
At the fluid-solid interface $\Gamma_{fs}$, we assume the continuity of the temperature field and the heat flux to adequately model the steady conjugate heat transfer between the fluid and solid:
\begin{equation}
\begin{aligned}
T_{f} &= T_{s} &\mbox{on}~~ \Gamma_{fs} \,, \\
- \boldsymbol{n}_{fs} \cdot k_{f} \nabla T_{f} &= - \boldsymbol{n}_{fs} \cdot k_{s}  \nabla T_{s} & \mbox{on}~~ \Gamma_{fs}. 
\end{aligned}
\label{eq: continuity of heat flux and temperature at fluid-solid interface}
\end{equation}
The normal $\boldsymbol{n}_{fs}$ at  $\Gamma_{fs}$ is here directed from the fluid into the solid domain.

As a result of the periodic distribution of solid material (\ref{eq: periodic solid material distribution}) in the channel, typically a (quasi-) periodically flow field will be established in the channel, provided that the flow remains steady and laminar, and the channel is sufficiently long.
At least, this requires that all other fluid properties, like the dynamic viscosity $\mu_f$ can be considered constant as well.
As shown in \citep{Buckinx2022arxiv, Buckinx2023arxiv}, the quasi-periodically developed field consists of a periodic velocity field $\boldsymbol{u}^{\star}_{f}$, on which an exponentially decaying mode with a periodic amplitude $\velamp_{f}$ is superposed:
\begin{equation}
\boldsymbol{u}_f(\boldsymbol{x}) =  \boldsymbol{u}^{\star}_{f}(\boldsymbol{x}) + \velamp_{f} (\boldsymbol{x}) \exp \left(\velocityeigenvalue \boldsymbol{\cdot} \boldsymbol{x} \right) \,,
\label{eq: quasi-periodically developed flow field }
\end{equation}
with $\boldsymbol{u}^{\star}_{f}(\boldsymbol{x}) = \boldsymbol{u}^{\star}_{f} \left(\boldsymbol{x} + \boldsymbol{l}_1 \right)$ and  $\velamp_{f}(\boldsymbol{x}) = \velamp_{f} \left(\boldsymbol{x} + \boldsymbol{l}_{1}\right)$.
The eigenvector $\velocityeigenvalue$ is parallel to the main flow direction $\boldsymbol{e}_1$, and its sign is such that $\velocityeigenvalue \boldsymbol{\cdot}\boldsymbol{e}_1 < 0 $.
Although the periodically developed flow field $\boldsymbol{u}^{\star}_{f}$ will also display transversal periodicity at a certain distance from the channel's side walls if the channel is wide enough, i.e.  $\boldsymbol{u}^{\star}_{f}(\boldsymbol{x}) = \boldsymbol{u}^{\star}_{f} \left(\boldsymbol{x} + \boldsymbol{l}_2 \right)$, the mode amplitude $\velamp_{f}$ will generally exhibit only streamwise periodicity along the periodicity vector $\boldsymbol{l}_1$.

The subject of the present work is to describe the temperature field $T$ according to the boundary-value problem (\ref{eq: problem statement: temperature field channel flow}) within the region of quasi-periodically developed flow, hence where the velocity field satisfies the asymptotic form (\ref{eq: quasi-periodically developed flow field }).

\section{Description of Periodically Developed Heat Transfer}
\label{sec: Description of Periodically Developed Heat Transfer}
If we substitute the first term of the asymptotic form for the velocity field (\ref{eq: quasi-periodically developed flow field }), i.e. $\boldsymbol{u}_f =  \boldsymbol{u}^{\star}_{f}$,  in the energy conservation equations (\ref{eq: problem statement: temperature field channel flow}), we retrieve the (partially) known solutions for temperature distribution in the periodically developed region, which we denote by $T_{\text{dev}}$.

\subsection{Constant Wall Temperature}
When a constant wall temperature $\bar{T}_s$ and  a zero heat flux apply to certain parts of the channel walls, as specified by the boundary conditions (\ref{eq: constant temperature and insulation walls}), the periodically developed temperature distribution is of the form
\begin{equation}
T_{\text{dev}}(\boldsymbol{x}) =  \bar{T}_s + \theta (\boldsymbol{x}) \exp \left(\temperatureeigenvalueconstantsolid \boldsymbol{\cdot} \boldsymbol{x} \right)\,,
\label{eq: temperature form constant solid temperature}
\end{equation}
as demonstrated in \citep{BuckinxBaelmans2015b}.
In this form, the mode amplitude $\theta$ is streamwise periodic, just like the developed flow field $ \boldsymbol{u}^{\star}_{f}$, so $\theta(\boldsymbol{x}) = \theta \left(\boldsymbol{x} + \boldsymbol{l}_1 \right)$.
If the developed flow field $ \boldsymbol{u}^{\star}_{f}$ is also transversally periodic, we have in addition $\theta(\boldsymbol{x}) = \theta \left(\boldsymbol{x} + \boldsymbol{l}_2 \right)$.
Further, the eigenvector $\temperatureeigenvalueconstantsolid$ has only a component along the main flow direction.
Consequently, the developed solution $T_{\text{dev}}$ for the boundary conditions (\ref{eq: constant temperature and insulation walls}) is completely determined by the following eigenvalue problem:
\begin{subequations}
	\label{eq: periodically developed heat transfer equations}
	\begin{align}
		\rho_{f} c_{f} \boldsymbol{u}^{\star}_{f} \boldsymbol{\cdot} \left(\nabla  \theta_{f}
		+ \boldsymbol{\lambda}_{T} \theta_f \right)
		&= \boldsymbol{\nabla} \boldsymbol{\cdot} k_{f} \boldsymbol{\nabla} \theta_f +  2 k_{f} \boldsymbol{\nabla} \theta_{f} \boldsymbol{\cdot} \temperatureeigenvalueconstantsolid + k_{f}\temperatureeigenvalueconstantsolid \boldsymbol{\cdot} \temperatureeigenvalueconstantsolid \theta_{f}
		\qquad \mbox{in } \Omega_{f} \,, 
		\label{eq: periodically developed fluid temperature theta equation}\\
		0
		&= \boldsymbol{\nabla} \boldsymbol{\cdot} k_{s} \boldsymbol{\nabla} \theta_{s} + 2  k_{s} \boldsymbol{\nabla} \theta_{s} \boldsymbol{\cdot} \temperatureeigenvalueconstantsolid + k_{s}\temperatureeigenvalueconstantsolid  \boldsymbol{\cdot} \temperatureeigenvalueconstantsolid \theta_{s}
		\qquad \mbox{in } \Omega_{s} \,, 
		\label{eq: periodically developed solid temperature theta  equation}\\
		\theta_{s}
		&=0
		\qquad \text{ on }  \Gamma_{D} 
		\label{eq: zero theta solid temperature  equation}\\
		- \boldsymbol{n}_{s} \boldsymbol{\cdot} k_s    \left( \boldsymbol{\nabla} \theta_s + \temperatureeigenvalueconstantsolid \theta_s   \right) 
		&= 0 
		\qquad \text{ on } \Gamma_{s} \setminus \Gamma_{D}  \,,
		\label{eq: zero flux channel wall theta temperature equation}  \\
		- \boldsymbol{n}_{fs} \boldsymbol{\cdot} k_{f}    \left( \boldsymbol{\nabla} \theta_{f} + \temperatureeigenvalueconstantsolid \theta_{f}  \right) 
		&= - \boldsymbol{n}_{fs} \boldsymbol{\cdot} k_{s}  \left( \boldsymbol{\nabla} \theta_{s}  + \temperatureeigenvalueconstantsolid  \theta_{s}   \right)   
		\qquad \text{on } \Gamma_{fs} \,, 
		\label{eq: fluid-solid interface continuous flux periodic temperature theta equation} \\
		\theta_{f}  
		&= \theta_{s} \qquad\text{ on } \Gamma_{fs} \,, 
		\label{eq: fluid-solid interface continuous temperature periodic temperature theta equation} \\
		\rho_f c_f \langle \boldsymbol{u}^{\star} \theta  \rangle \boldsymbol{\cdot} \temperatureeigenvalueconstantsolid
		&= \langle k \boldsymbol{\nabla}  \theta  \rangle \boldsymbol{\cdot} \temperatureeigenvalueconstantsolid + \langle k \theta \rangle \temperatureeigenvalueconstantsolid  \boldsymbol{\cdot} \temperatureeigenvalueconstantsolid  \,,
		\label{eq: eigenvalue equation}  \\
		\langle \boldsymbol{u}^{\star} \rangle \boldsymbol{\times} \temperatureeigenvalueconstantsolid &=0\, , 
		\label{eq: eigenvalue vector parallel to flow direction} \\
		\theta_{f} \left( \boldsymbol{x}  \right) 
		&= \theta_{f}  \left( \boldsymbol{x} +  \boldsymbol{l}_{j}  \right) \,, 
		\label{eq: periodicity theta fluid  equation}\\
		\theta_{s} \left( \boldsymbol{x}  \right) 
		&= \theta_{s}  \left( \boldsymbol{x} +  \boldsymbol{l}_{j}  \right) \,,
		\label{eq: periodicity theta solid  equation} \\
		\langle \theta \rangle &= \theta_0 \,.
		\label{eq: volume-average theta} 
	\end{align}
	\label{eq: eigenvalue problem theta}
\end{subequations}
This quadratic eigenvalue problem has been obtained by substituting the form (\ref{eq: temperature form constant solid temperature}) into the original energy conservation equations (\ref{eq: problem statement: temperature field channel flow}) and boundary conditions (\ref{eq: constant temperature and insulation walls}) and (\ref{eq: continuity of heat flux and temperature at fluid-solid interface}).
In the specific case where the entire solid temperature is uniform, i.e. $T_s=\bar{T}_{s}$, it reduces to the eigenvalue problem presented in \citep{BuckinxBaelmans2015b}.
As such, the present problem is a generalization of the former towards the situation where conjugate heat transfer occurs.
Based on physical considerations, of all its possible solutions $(\temperatureeigenvalueconstantsolid, \theta)$ only the one whose eigenvector has the smallest real negative component along the main flow direction is of interest, so $\temperatureeigenvalueconstantsolid \boldsymbol{\cdot} \boldsymbol{e}_1  < 0$ \citep{BuckinxBaelmans2015b}.

Due to the streamwise periodicity conditions (\ref{eq: periodicity theta fluid  equation}), the eigenvalue problem (\ref{eq: temperature form constant solid temperature}) needs to be solved on just a single row of the array.
When there is also transversal periodicity, it can even be solved on a single unit cell of the array.
Therefore, the overall energy balance (\ref{eq: eigenvalue equation}) has been obtained by volume-averaging the energy conservation laws (\ref{eq: problem statement: temperature field channel flow}) over the row (or unit cell) of the array, via the volume-averaging operator $\langle \; \rangle$.
We clarify that the thermal conductivity $k$ which appears within the volume-averaging operator  $\langle \; \rangle$ in (\ref{eq: eigenvalue equation}), should be interpreted as a distribution, so $k=k_f$ in $\Omega_f$ and $k=k_s$ in $\Omega_s$.
Further, we have assumed that the main flow direction is given by the direction of the volume-averaged velocity distribution $\langle \boldsymbol{u}^{\star}\rangle$ according to equation (\ref{eq: eigenvalue vector parallel to flow direction}), thus $\langle \boldsymbol{u}^{\star}\rangle = \Vert\langle \boldsymbol{u}^{\star} \rangle \Vert \boldsymbol{e}_1 $.
In order to find a unique solution for the mode amplitude $\theta$, its volume-averaged value $\theta_0$ over the row (or unit cell) of the array has been imposed, in accordance with (\ref{eq: volume-average theta}). 
In practice, this value is not known since it depends on the specific inlet conditions (\ref{eq: inlet outlet bcs temperature}).
Hence, it must be estimated or modelled if the absolute temperature level in the channel is of interest.

\subsection{Constant Wall Heat Flux}
When a constant heat flux $q_s$ and a zero heat flux apply to certain parts of the channel walls, as specified by the boundary conditions (\ref{eq: constant heat flux  and insulation walls}), the periodically developed temperature distribution $T_{\text{dev}}$ has a different form:
\begin{equation}
\label{eq: linear-periodic temperature constant flux}
T_{\text{dev}}(\boldsymbol{x}) = \boldsymbol{\nabla} \mathrm{T}_{\text{dev}} \boldsymbol{\cdot} \boldsymbol{x} + T^{\star}(\boldsymbol{x})\,,
\end{equation}
as shown in \citep{BuckinxBaelmans2016}.
The former solution consists of a first contribution which varies linearly over space with a constant temperature gradient $\boldsymbol{\nabla} \mathrm{T}_{\text{dev}}$ along the main flow direction, and a second contribution $T^{\star}$ which inherits the streamwise (and transversal) periodicity of the developed flow field:
$T^{\star} (\boldsymbol{x}) = T^{\star} (\boldsymbol{x}+\boldsymbol{l}_j)$.
Both contributions are governed by the following periodic temperature problem, which can be solved on a row or unit cell of the array, depending on whether the former periodicity condition for $ T^{\star}$ holds for $j=1$ or $j\in\{1,2\}$:
\begin{subequations}
\label{eq: periodically developed heat transfer equations constant flux}
\begin{align}
\rho_{f} c_{f} \boldsymbol{u}^{\star}_{f} \boldsymbol{\cdot} \left(\nabla  T^{\star}_{f} 
+ \mathrm{\boldsymbol{\nabla} T}_{\text{dev}} \right) &= \boldsymbol{\nabla} \boldsymbol{\cdot} k_f \boldsymbol{\nabla} T^{\star}_{f} 
\qquad \mbox{in } \Omega_{f} \,, 
\label{eq: periodically developed temperature equation fluid constant flux}\\
0
&= \boldsymbol{\nabla} \boldsymbol{\cdot} k_{s} \boldsymbol{\nabla} T^{\star}_{s}  
\qquad \mbox{in } \Omega_{s} \,, 
\label{eq: periodically developed temperature equation solid  constant flux}\\
T^{\star} \left( \boldsymbol{x}  \right) &= T^{\star} \left( \boldsymbol{x} +  \boldsymbol{l}_{j}  \right)\,, \\
- \boldsymbol{n} \boldsymbol{\cdot} k_{s} \left( \nabla T^{\star}_s + \mathrm{\boldsymbol{\nabla} T}_{\text{dev}} \right) &= q_{s}  \qquad \text{on } \Gamma_{N}  \,,
\label{eq: Neumann bc periodic temperature equation constant flux} \\
- \boldsymbol{n} \boldsymbol{\cdot} k_s  \left(  \nabla T^{\star}_s + \mathrm{\boldsymbol{\nabla} T}_{\text{dev}} \right) &= 0  \qquad \text{ on } \Gamma_{s} \setminus \Gamma_{N}  \,,
\label{eq: zero Neumann bc periodic temperature equation constant flux} \\
\rho_{f} c_{f} \langle \boldsymbol{u}^{\star} \rangle \boldsymbol{\cdot} \mathrm{\boldsymbol{\nabla} T}_{\text{dev}} &= \langle q_{s} \delta_N \rangle \,,
\label{eq: average temperature gradient equation constant flux}  \\
\langle \boldsymbol{u}^{\star} \rangle \boldsymbol{\times} \mathrm{\boldsymbol{\nabla} T}_{\text{dev}} &=0\,,
\label{eq: average temperature gradient parallel to flow direction}  \\
\langle T^{\star}\rangle &= T^{\star}_{0} \,. 
\label{eq: volume-average periodic temperature unit cell} 
\end{align}
\end{subequations}
This periodic temperature problem follows straightforward from the original energy conservation equations (\ref{eq: problem statement: temperature field channel flow}) and boundary conditions (\ref{eq: constant temperature and insulation walls}) and (\ref{eq: continuity of heat flux and temperature at fluid-solid interface}), after substitution of (\ref{eq: linear-periodic temperature constant flux}).
In absence of channel walls, it becomes equivalent to the periodic temperature problem presented in \citep{BuckinxBaelmans2016}.
We have assumed that the direction of the mean temperature gradient $\mathrm{\boldsymbol{\nabla} T}_{\text{dev}}$ is parallel to the main flow direction as well the direction of the volume-averaged velocity over the row (or unit cell) of the array.
To formulate the overall energy balance over the row (or unit cell) of the array (\ref{eq: average temperature gradient equation constant flux}), we have relied on the Dirac indicator $\delta_N$ associated with the surface $\Gamma_N$, as defined in the works by \citet{Schwartz1966, Gagnon1970}.
Again, the constant $T^{\star}_{0}$ which determines the volume-averaged value of the periodic temperature part $T^{\star}$, is not known in practice.
Therefore, it must be modeled somehow, if the absolute temperature level in the channel is of interest.

\section{Description of Quasi-Periodically Developed Heat Transfer}
\label{sec: Description of Quasi-Periodically Developed Heat Transfer}
To construct a solution for the temperature distribution in the region where the flow is quasi-periodically developed, we propose the following
ansatz
\begin{equation}
\label{eq: quasi-periodic exponential ansatz}
T = T_{\text{dev}} + \Theta \exp \left(\temperatureeigenvaluequasideveloped \boldsymbol{\cdot} \boldsymbol{x} \right) 
\end{equation}
with $\Theta \ll  T_{\text{dev}}$.
This means that we restrict our description to small (exponentially decaying) perturbations upstream or downstream of the periodically developed region, in analogy to the asymptotic velocity form (\ref{eq: quasi-periodically developed flow field }).

When the constant-wall-temperature conditions (\ref{eq: constant temperature and insulation walls}) apply, we then find that the amplitude and decay rate $(\Theta, \temperatureeigenvaluequasideveloped)$ of the former temperature perturbation must satisfy the same eigenvalue problem as the solution pair $(\theta, \temperatureeigenvalueconstantsolid)$ to (\ref{eq: temperature form constant solid temperature}):
\begin{subequations}
	\begin{align}
		\rho_{f} c_{f} \boldsymbol{u}^{\star}_{f} \boldsymbol{\cdot} \left(\nabla  \Theta_{f}
		+ \temperatureeigenvaluequasideveloped \Theta_f \right)
		&= \boldsymbol{\nabla} \boldsymbol{\cdot} k_{f} \boldsymbol{\nabla} \Theta_f +  2 k_{f} \boldsymbol{\nabla} \Theta_{f} \boldsymbol{\cdot} \temperatureeigenvaluequasideveloped + k_{f}\temperatureeigenvaluequasideveloped \boldsymbol{\cdot} \temperatureeigenvaluequasideveloped  \Theta_{f}
		\quad \mbox{in } \Omega_{f} \,, 
		\label{eq: periodically developed fluid temperature quasi-developed equation}\\
		0
		&= \boldsymbol{\nabla} \boldsymbol{\cdot} k_{s} \boldsymbol{\nabla} \Theta_{s} + 2  k_{s} \boldsymbol{\nabla} \Theta_{s} \boldsymbol{\cdot} \temperatureeigenvaluequasideveloped + k_{s}\temperatureeigenvaluequasideveloped \boldsymbol{\cdot} \temperatureeigenvaluequasideveloped \Theta_{s}
		\quad \mbox{in } \Omega_{s} \,, 
		\label{eq: periodically developed solid temperature quasi-developed  equation}\\
		\Theta_{s}
		&=0
		\qquad \text{ on }  \Gamma_{D} 
		\label{eq: zero quasi-developed solid temperature  equation}\\
		- \boldsymbol{n}_{s} \boldsymbol{\cdot} k_s    \left( \boldsymbol{\nabla} \Theta_s + \temperatureeigenvaluequasideveloped \Theta_s   \right) 
		&= 0 
		\qquad \text{ on } \Gamma_{s} \setminus \Gamma_{D}  \,,
		\label{eq: zero flux channel wall quasi-developed temperature equation}  \\
		- \boldsymbol{n}_{fs} \boldsymbol{\cdot} k_{f}    \left( \boldsymbol{\nabla} \Theta_{f} + \temperatureeigenvaluequasideveloped \Theta_{f}  \right) 
		&= - \boldsymbol{n}_{fs} \boldsymbol{\cdot} k_{s}  \left( \boldsymbol{\nabla} \Theta_{s}  + \temperatureeigenvaluequasideveloped \Theta_{s}   \right)   
		\qquad \text{on } \Gamma_{fs} \,, 
		\label{eq: fluid-solid interface continuous flux periodic temperature quasi-developed equation} \\
		\Theta_{f}  
		&= \Theta_{s} \qquad\text{ on } \Gamma_{fs} \,, 
		\label{eq: fluid-solid interface continuous temperature periodic temperature quasi-developed equation} \\
		\rho_f c_f \langle \boldsymbol{u}^{\star} \Theta \rangle \boldsymbol{\cdot} \temperatureeigenvaluequasideveloped
		&= \langle k \boldsymbol{\nabla}  \Theta \rangle \boldsymbol{\cdot} \temperatureeigenvaluequasideveloped + \langle k \Theta \rangle \temperatureeigenvaluequasideveloped \boldsymbol{\cdot} \temperatureeigenvaluequasideveloped  \,,
		\label{eq: eigenvalue equation quasi-developed}  \\
		\langle \boldsymbol{u}^{\star} \rangle \boldsymbol{\times} \temperatureeigenvaluequasideveloped &=0\, , 
		\label{eq: eigenvalue vector parallel to flow direction quasi-developed} \\
		\Theta_{f} \left( \boldsymbol{x}  \right) 
		&= \Theta_{f}  \left( \boldsymbol{x} +  \boldsymbol{l}_{j}  \right) \,, 
		\label{eq: periodicity quasi-developed fluid  equation}\\
		\Theta_{s} \left( \boldsymbol{x}  \right) 
		&= \Theta_{s}  \left( \boldsymbol{x} +  \boldsymbol{l}_{j}  \right) \,,
		\label{eq: periodicity quasi-developed solid  equation} \\
		\langle \Theta \rangle &= \Theta_0 \,.
		\label{eq: volume-average quasi-developed} 
	\end{align}
	\label{eq: eigenvalue problem quasi-developed}
\end{subequations}
This follows directly from the linearity of the energy conservation equations (\ref{eq: temperature form constant solid temperature}) and the fact that we may neglect the advection of thermal energy by the small velocity mode to obtain an asymptotic approximation for the temperature distribution: $\velamp_{f} \exp \left(\velocityeigenvalue \boldsymbol{\cdot} \boldsymbol{x} \right) \boldsymbol{\cdot} \boldsymbol{\nabla} T_{\text{dev}} \ll \boldsymbol{u}^{\star}_f  \boldsymbol{\cdot} \boldsymbol{\nabla} \left(\Theta \exp \left(\temperatureeigenvaluequasideveloped \boldsymbol{\cdot} \boldsymbol{x} \right)  \right)$.
The implication is that $\temperatureeigenvaluequasideveloped$ thus corresponds to the eigenvector solution to (\ref{eq: temperature form constant solid temperature}) whose magnitude is second smallest after $ \temperatureeigenvalueconstantsolid$.

In general, the mode amplitude $\Theta$ in the quasi-developed region will display only streamwise periodicity, even where the velocity field $\boldsymbol{u}^{\star}_{f}$ is both streamwise and transversally periodic.
Therefore, the eigenvalue problem for $(\Theta, \temperatureeigenvaluequasideveloped)$ should be solved on an entire row of the array, for some prescribed temperature perturbation size $ \Theta_0 $.
As the latter depends again on the specific inlet conditions for the velocity and temperature field in the channel, we can in practice reconstruct the temperature field in the quasi-developed region only up to the two unknown constants $ \theta_0 $ and $ \Theta_0 $.

Also when the constant-heat-flux conditions (\ref{eq: constant heat flux and insulation walls}) apply, the temperature mode and its eigenvalue $(\Theta, \temperatureeigenvaluequasideveloped)$ in the quasi-developed region will satisfy nearly the same eigenvalue problem (\ref{eq: temperature form constant solid temperature}) as the temperature modes for an isothermal wall.
However, the Dirichlet boundary condition (\ref{eq: zero quasi-developed solid temperature  equation}) should be replaced by a Neumann boundary condition.
So instead of the boundary conditions (\ref{eq: zero quasi-developed  solid temperature  equation}) and (\ref{eq: zero flux channel wall quasi-developed  temperature equation}), we then have 
\begin{align}
- \boldsymbol{n}_{s} \boldsymbol{\cdot} k_s    \left( \boldsymbol{\nabla}  \Theta_s + \temperatureeigenvaluequasideveloped  \Theta_s   \right) 
&= 0 
\qquad \text{ on } \Gamma_{s}   \,.
\label{eq: periodically quasi-developed heat transfer constant flux Neumann bcs}
\end{align}

\section{Discussion}
At this point, the eigenvalue problem (\ref{eq: eigenvalue problem quasi-developed}) only tells us what kind of exponential temperature perturbations, or modes, are allowed by the energy conservation equations (\ref{eq: temperature form constant solid temperature}).
In itself, it does not give any indication whether these allowed modes will effectively occur, and if so, how important they are.
After all, the underlying ansatz (\ref{eq: quasi-periodic exponential ansatz}) might be only valid over a very short distance upstream (or downstream) of the periodically developed region, depending on the actual inlet and outlet conditions (\ref{eq: inlet outlet bcs temperature}).
For that reason, the onset of the quasi-developed temperature modes (\ref{eq: quasi-periodic exponential ansatz}) requires further empirical evidence by performing full-scale (DNS) simulations of the developing temperature and flow fields.
In our subsequent work \citep{Vangeffelen2024}, such empirical evidence will be provided in particular
for mini and micro channels channels containg arrays of periodic offset-strip fins.
There,  we will also show that the eigenvalue solutions to (\ref{eq: eigenvalue problem quasi-developed}) can explain the validity of the existing macro-scale models for developed heat transfer, at least for a constant-heat-flux condition.
In the near future, the present work will be extended by including a numerical study on the onset of the quasi-periodically developed heat transfer regimes in channels with arrays of square cylinders, as considered in the works \citep{Buckinx2022arxiv,Buckinx2023arxiv}.

\section{Acknowledgements}
\label{sec:acknowledgements}
The work presented in this paper was partly funded by the Research Foundation — Flanders (FWO) through the post-doctoral project grant 12Y2919N of G. Buckinx.

\clearpage

\bibliographystyle{abbrvnat}
\setcitestyle{authoryear,open={((},close={))}} 

\bibliography{References}
	
\end{document}